\documentclass{article}
\usepackage{amsmath}
\usepackage{graphicx}
\title{\textbf{Colored flux tube in the Euclidean spacetime}}

\date{}

\begin{document}
\maketitle

\hspace{2cm}
\parbox{0.7\linewidth}{
\textbf{\Large{Vladimir Dzhunushaliev}} \\
\vspace{6pt}

\textit{
Institut f\"ur Mathematik, Universit\"at Potsdam\\ 
D-14469, Potsdam, Germany \\
and \\
Dept. Phys. and Microel. Engineer. \\
Kyrgyz-Russian Slavic University\\
Kievskaya Str. 44, 720021, Bishkek, Kyrgyz Republic\\
\textit{E-mail: dzhun@hotmail.kg}}
}

\vspace{36pt}
The flux tube solution in the Euclidean spacetime with the color longitudinal 
electric field in the SU(2) Yang - Mills - Higgs theory with broken gauge 
symmetry is found. Some arguments are given that this flux tube is a pure 
quantum object in the SU(3) quantum theory reduced to the SU(2) 
Yang - Mills - Higgs theory. 
\par
\vspace{12pt}
{\flushleft Key words : flux tube, broken gauge symmetry, nonperturbative quantization}

\vspace{24pt}
\section{INTRODUCTION}
\vspace{12pt}

The confinement problem in quantum chromodynamics (QCD) 
is naturally connected with the existence of a hypothesized 
flux tube filled with a color longitudinal electric field and stretched 
between quark and antiquark. Such tube confines quark and 
antiquark in a pair and does not give any possibility to destroy 
such pair. Evidently the derivation of the colored non-Abelian flux tube 
is impossible in classical gauge field theory\footnote{altough in dual 
theories exist the well known Nilesen-Olesen flux tube solutions \cite{no} 
filled with a longitudinal Abelian magnetic field $H_z$.} 
and still remains an unresolved 
problem in QCD. In this letter we would like to show that the 
flux tube solution exists in the SU(2) Yang - Mills - Higgs 
theory with broken gauge symmetry. The gauge symmetry breakdown 
is a quantum phenomenon therefore we are searching for the flux tube 
solution in the classical theory with some additional quantum term. 
It means that we consider an approximate model of the flux tube 
with the color electric field. At the end of this letter we will 
argue that the above mentioned Yang - Mills - Higgs theory can be 
obtained as an approximation of quantum SU(3) gauge theory where 
the SU(3) gauge potential is splitted on two parts: the first 
one is the SU(2) gauge potential (SU(2) is a small subgroup of SU(3): 
$SU(2) \in SU(3)$) and the second one is the coset $SU(3)/SU(2)$. 
The SU(2) degrees of freedom remain almost classical but the coset 
degrees of freedom are completely quantum and can be presented as 
a scalar field (physically it is a condensate of the coset fields). 
In this presentation SU(2) and coset fields are similar to a magnetic 
field and wave function of Cooper pairs in a superconductor. 

\vspace{24pt}
\section{INITIAL EQUATIONS}
\vspace{12pt}

Let us start from the Yang - Mills - Higgs field equations in the Euclidean 
spacetime with broken gauge symmetry
\begin{eqnarray}
  \mathcal{D}_\nu F^{a\mu\nu} &=& g \epsilon^{abc} \phi^b 
  \mathcal{D}^\mu \phi^c - m^2 A^{a\mu} , 
\label{sec1-10}\\
  \mathcal{D}_\mu \mathcal{D}^\mu \phi^a &=& -\lambda \phi^a 
  \left(
  \phi^b \phi^b - \phi^2_\infty
  \right) 
\label{sec1-20}
\end{eqnarray}
here $F_{a\mu\nu} = \partial_\mu A^a_\nu - \partial_\nu A^a_\mu 
+ g \epsilon^{abc} A^b_\mu A^c_\nu$ is the field tensor for the SU(2) gauge 
potential $A^a_\mu$; $a,b,c = 1,2,3$ are the color indices; 
$D_\nu [\cdots ]^a = \partial_\nu [\cdots ]^a + 
g \epsilon^{abc} A^b_\mu  [\cdots ]^c$ is the gauge derivative; $\phi^a$ 
is the Higgs field; $g$ and $\phi^a_\infty$ are some constants; 
$m^2 A^a_\mu$ is the most important term for us which destroys the gauge 
invariance of the Yang - Mills - Higgs theory. 
\par
The solution we search in the following form 
\begin{equation}
    A^1_\tau = \frac{f(\rho)}{g} , \quad A^2_z = \frac{v(\rho)}{g} , 
    \quad \phi^3 = \frac{\phi(\rho)}{g} 
\label{sec1-25}
\end{equation}
here $z, \rho , \varphi$ are cylindrical coordinate system, $\tau$ is the 
Euclidean time. The substitution to the Euclidean Yang - Mills - Higgs 
equations gives us 
\begin{eqnarray}
    f'' + \frac{f'}{\rho} &=& f \left( \phi^2 + v^2 - m^2 \right),
\label{sec1-30}\\
    v'' + \frac{v'}{\rho} &=& v \left( \phi^2 + f^2 - m^2 \right),
\label{sec1-40}\\  
    \phi'' + \frac{\phi'}{\rho} &=& \phi \left[ f^2 + v^2 
    + \lambda \left( \phi^2 - \phi^2_\infty \right)\right]
\label{sec1-50}
\end{eqnarray}
here we redefined $g \phi \rightarrow \phi$, $g f \rightarrow f$, 
$g v \rightarrow v$ and $\lambda/g^2 \rightarrow \lambda$. 
The similar equations in the Lorentzian spacetime 
with the presence of $A^a_\varphi$ (and consequantley 
with a magnetic field $H^a_z$) was investigated in Ref. \cite{dzhun} and 
it was shown that exist: (1) the flux tube filled with electric/magnetic fields 
on the background of an external constant magnetic/electric field; (2) 
the Nielsen-Olesen flux tube dressed with transversal color electric 
and magnetic fields. 
\par 
We will consider the simplest case $f = v$. Then we have 
\begin{eqnarray}
    f'' + \frac{f'}{x} &=& f \left( \phi^2 + f^2 - m^2 \right),
\label{sec1-60}\\
    \phi'' + \frac{\phi'}{x} &=& \phi \left[ 2 f^2 + 
    \lambda \left( \phi^2 - \phi^2_\infty \right)\right]
\label{sec1-70}
\end{eqnarray}
here we redefined $f/\alpha \rightarrow f$, 
$\phi /\alpha \rightarrow \phi$ and $m/\alpha \rightarrow m$, 
$\phi_\infty /\alpha \rightarrow \phi_\infty$ and 
$\rho \alpha \rightarrow x$; $\alpha$ is some constant which will be 
defined later. 

\vspace{24pt}
\section{NUMERICAL INVESTIGATION}
\vspace{12pt}

We will solve Eq(s). \eqref{sec1-60} \eqref{sec1-70} with an 
iterative procedure. On the $i$ step Eq. \eqref{sec1-60} has 
the following form 
\begin{equation}
    f_i'' + \frac{f_i'}{x} = f_i \left( \phi^2_{i-1} + f^2_i - m^2_i \right),
\label{sec2-10}\\
\end{equation}
here $\phi_{i-1}$ was defined on $(i-1)$ step and the null approximation is 
$\phi_0 = 2 - 1/\cosh^2 (x/4)$. The numerical investigation for Eq. 
\eqref{sec2-10} shows that there is a number $m^*_i$ for which: for 
$m_i > m^*_i$ the solution is singular $f_i(x) \rightarrow - \infty$ 
by $x \rightarrow x_0$ and for $m_i < m^*_i$ the solution is also singular $f_i(x) \rightarrow + \infty$ by $x \rightarrow x_0$. It means 
that for the value $m_i = m^*_i$ the solution $f(x)$ is regular one. The value 
$m^*_i$ can be defined with the method of iterative approximation. 
\par
The next step on the $i$ iteration is solving of Eq. \eqref{sec1-70}
\begin{equation}
    \phi_i'' + \frac{\phi_i'}{x} = \phi_i \left[ 2 f^2_i + 
    \lambda \left( \phi^2_i - \phi^2_{i \infty} \right)\right].
\label{sec2-20}
\end{equation}
The numerical investigation for this equation shows that there is a number 
$\phi_{i \infty}$ for which: for $\phi_{i \infty} > \phi^*_{i \infty}$ 
the function $\phi_i (x)$ oscillates with decreasing amplitude, for 
$\phi_{i \infty} < \phi^*_{i \infty}$ the function 
$\phi_i (x) \rightarrow + \infty$ by $x \rightarrow x_0$. The number 
$\phi^*_i$ defines a regular solution $\phi_i(x)$. 
\par
The value $f(0)$ can be arbitrary but we choose them by such a way that 
$f(0)/\alpha = 0.5$. Thus in Eq(s) \eqref{sec1-60} \eqref{sec1-70} 
we have only two independent parameters $\lambda$ and $\phi(0)$. In the 
calculations presented here we take $\lambda = 0.1$ and 
$\phi(0)/\alpha = 1.0$. 
\par
\begin{figure}[h]
    \begin{center}
    \fbox{
        \includegraphics[width=10cm,height=7cm]{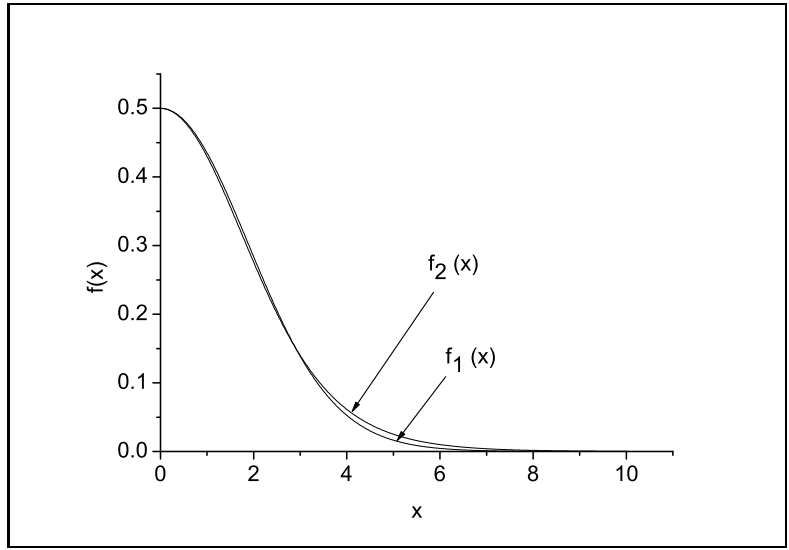}}
        \caption{The functions $f_1(x)$ and $f_2(x)$. The first and second 
        iterations practically coincide.}
        \label{fig1}
    \end{center}
\end{figure}
\begin{figure}[h]
    \begin{center}
    \fbox{
        \includegraphics[width=10cm,height=7cm]{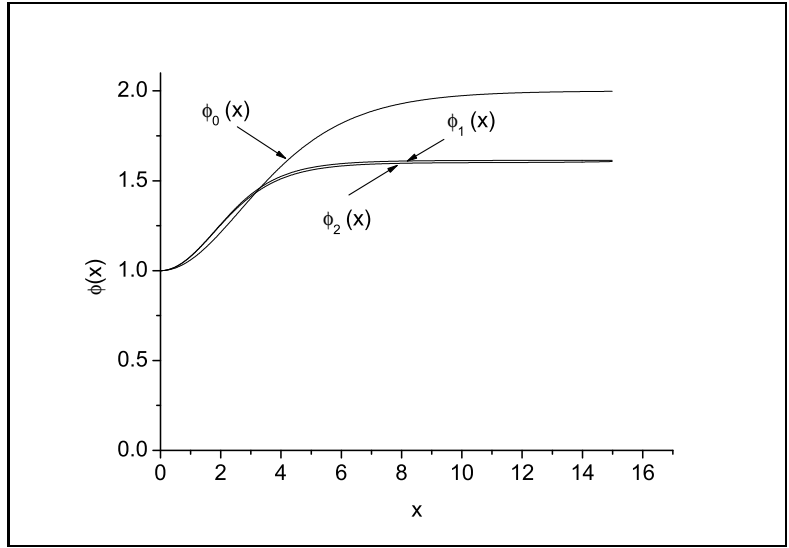}}
        \caption{The functions $\phi_0(x)$, $\phi_1(x)$ and $\phi_2(x)$. 
        The first and second iterations practically coincide.}
        \label{fig2}
    \end{center}
\end{figure}

The iterative process described above gives us the $m^*_i$ and 
$\phi^*_i$ presented on Table \ref{table1}. 
\begin{table}[h]
    \begin{center}
        \begin{tabular}{|c|c|c|c|c|}\hline
          i & 1 & 2& 3 & 4 \\ \hline
            $m^*_i$ & 1.342794\ldots & 1.362562\ldots & 1.359847\ldots 
            & 1.359851\ldots \\ \hline
            $\phi^*_i$ & 1.613484\ldots & 1.601260\ldots & 1.606675\ldots 
            & 1.606186\ldots \\ \hline
        \end{tabular}
    \end{center}
    \caption{The iterative parameters $m^*_i$ and $\phi^*_i$.}
    \label{table1}
\end{table}
The functions $f_i(x)$ and $\phi_i(x)$ for $i=1,2$ are presented 
on Fig's \ref{fig1}, \ref{fig2}. 
\begin{figure}[h]
    \begin{center}
    \fbox{
        \includegraphics[width=10cm,height=7cm]{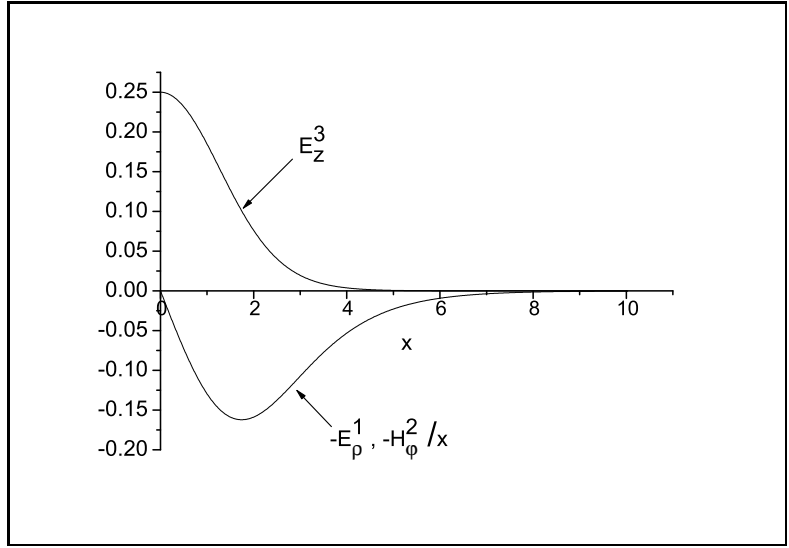}}
    \end{center}
\caption{The electric fields $E^3_z(x)$, $E^1_\rho(x)$ and magnetic field 
$H^2_\varphi(x) = - E^1_\rho(x)$}
\label{fig3}
\end{figure}
\par 
The electric and magnetic fields are 
\begin{eqnarray}
    E^3_z(x) &=& f(x) v(x) = f^2(x) ,
\label{sec2-30}\\
  E^1_\rho(x) &=& f'(x) ,
\label{sec2-40}\\
  H^2_\varphi (x) &=& x \epsilon_{\varphi \rho z} F^{\rho z} = -xf'(x)
\label{sec2-45}
\end{eqnarray}
These fields are presented on Fig. \ref{fig3}. 
\par 
It is easy to show that the asymptotical behaviour of 
the regular solutions $f^*(x)$ and $\phi^*(x)$ is 
\begin{eqnarray}
    f^*(x) &=& f_0 \frac{e^{-x\sqrt{\phi^{*2}_\infty - m^{*2}}}}{\sqrt{x}} + 
    \cdots ,
\label{sec2-50}\\
  \phi^*(x) &=& \phi^*_\infty - 
  \phi_0 \frac{e^{-x\sqrt{2\lambda \phi^{*2}_\infty}}}{\sqrt{x}} + \cdots 
\label{sec2-60}
\end{eqnarray}
here $f_0$ and $\phi_0$ are some constants. 

\vspace{24pt}
\section{QUANTUM INTERPRETATION OF \\ OBTAINED COLORED FLUX TUBE}
\vspace{12pt}

In Ref. \cite{vdsin2} it is shown that having some assumptions about 
the non-perturbative 
2 and 4-points Greens functions in the SU(3) quantum Yang - Mills theory 
it can be reduced to the classical SU(2) Yang - Mills - Higgs 
theory plus some extra term which is zero for the ansatz \eqref{sec1-25}. 
In short this $SU(3) \rightarrow SU(2)$ reduction can be presented by the 
following manner (here we follow to Ref. \cite{vdsin2}). At first we decompose 
the SU(3) gauge potential on ordered and disordered phases: 
\begin{enumerate}
\item The gauge field components $A^a_\mu \in SU(2), a=1,2,3$ 
      belonging to the small subgroup $SU(2) \subset SU(3)$ are in an ordered phase. 
      It means that 
\begin{equation}
  \left\langle A^a_\mu (x) \right\rangle  \approx (A^a _{\mu} (x))_{cl}.
\label{sec3-10}
\end{equation}
      The subscript means that this is the classical field. It is assumed 
      that in the first approximation these degrees of freedom are classical 
      and is described by SU(2) Yang - Mills equations. 
      $\left\langle \ldots \right\rangle$ is a quantum average. 
\item The gauge field components $A^m_\mu$ (m=4,5, ... , 8) and 
      $A^m_\mu \in SU(3)/SU(2)$) belonging to the coset SU(3)/SU(2) are in 
      a disordered phase (or in other words, they are condensed). It means that 
\begin{equation}
  \left\langle A^m_\mu (x) \right\rangle = 0, 
  \quad \text{but} \quad 
  \left\langle A^m_\mu (x) A^n_\nu (x) \right\rangle \neq 0 .
\label{sec3-20}
\end{equation}
      These degrees of freedom are pure quantum degrees and are involved in the 
      equations for the ordered phase as an averaged field distribution of coset 
      components. 
\end{enumerate}
In Ref. \cite{vdsin2} was made the following assumptions and simplifications:
\begin{enumerate}
\item 
    The correlation between coset components $A^m_\mu (y)$ and $A^n_\nu (x)$ 
    in two points $x^\mu$ and $y^\mu$ is 
    \begin{equation}
        \left\langle A^m_\mu (y) A^n_\nu (x) \right\rangle =
        - \eta_{\mu\nu} \mathcal{G}^{mn} (y,x).
    \label{sec3-30}
    \end{equation}
    The function $\mathcal{G}^{mn} (y,x)$ can be presented in one function 
    approximation as 
    \begin{equation}
        \mathcal{G}^{mn} (y,x) = - \frac{1}{3}f^{mpb} f^{npc} 
        \phi^b (y) \phi^c (x) 
    \label{sec3-40}    
    \end{equation}
    where $f^{abc}$ is the structural constants of the SU(3) group. 
\item There is no correlation between 
      ordered (classical) and disordered (quantum) phases 
    \begin{equation}
      \left\langle f(a^a_\mu) g(A^m_\nu) \right\rangle =
      f(a^a_\mu)  \left\langle g(A^m_\mu) \right\rangle
    \label{sec3-50}
    \end{equation}
    where $f$ and $g$ are arbitrary functions. 
\item 
    In the first approximation the 4-point Green's function is 
    \begin{equation}
    \begin{split}
        &\left\langle
        A^m_\alpha (x) A^n_\beta (y) A^p_\mu (z) A^q_\nu (u) 
        \right\rangle = 
        \\
        &\left(
        E^{mnpq}_{1,abcd} \eta_{\alpha\beta} \eta_{\mu\nu} + 
        E^{mpnq}_{2,abcd} \eta_{\alpha\mu} \eta_{\beta\nu} + 
        E^{mqnp}_{3,abcd} \eta_{\alpha\nu} \eta_{\beta\mu}
        \right) 
        \\
        &\phi^a (x) \phi^b(y) \phi^c (z) \phi^d(u) 
\label{sec3-60}
\end{split}
\end{equation}
\end{enumerate} 
here $E^{mnpq}_{1,abcd}, E^{mpnq}_{2,abcd}, E^{mqnp}_{3,abcd}$ are some constants. 
The main idea proposed in Ref. \cite{vdsin2} is that the initial SU(3) Lagrangian 
\begin{equation}
    \mathcal{L}_{SU(3)} = -\frac{1}{4}F^A_{\mu\nu} F^{A\mu\nu}, \, 
    A = 1,2, \cdots 8 .
\label{sec3-70}
\end{equation}
after above mentioned assumptions and simplifications can be reduced to the 
SU(2) Yang - Mills - Higgs Lagrangian 
\begin{equation}
\begin{split}
  \mathcal{L}_{SU(2)} = &- \frac{1}{4}  F^a_{\mu\nu} F^{a\mu\nu} + 
  \frac{1}{2} \left(
  \partial_\mu \phi^a + \frac{g}{2} \epsilon^{abc} A^b_\mu \phi^c
  \right)^2 + 
  \\
  &\frac{m^2_\phi}{2} (\phi ^a \phi ^a ) 
  - \lambda \left( \phi^a \phi^a \right)^2  +
  \frac{g^2}{2} a_{\mu} ^b \phi ^b a^{c \mu} \phi ^c .
\label{sec3-80}
\end{split}
\end{equation}
Here is also assumed that there is the gauge symmetry breakdown lieading to 
the term $\frac{m^2_\phi}{2} (\phi ^a \phi ^a )$. 
The first term $F^a_{\mu\nu} F^{a\mu\nu}$ is the Lagrangian for the 
ordered phase $A^a_\mu$ (the SU(2) Lagrangian) and the next 3 terms are 
the Lagrangian for the disordered (condensate) phase (Higgs Lagrangian). 
There is an additional gauge noninvariant term 
$\frac{g^2}{2} a_{\mu} ^b \phi ^b a^{c \mu} \phi ^c$. For the ans\"atz 
\eqref{sec1-25} the corresponding terms in field equations are zero and 
it is unimportant for us. 
\par
Finally, in the context of offered here $SU(3) \rightarrow SU(2)$ reduction 
the colored flux tube obtained above is \textit{a pure quantum object in the 
SU(3) Yang - Mills theory}. 

\vspace{24pt}
\section{CONCLUSIONS}
\vspace{12pt}

In this letter we have shown that in the SU(2) Yang - Mills - Higgs theory 
can exist the flux tube solutions filled with the color longitudinal electric 
field for which it is necessary to have a gauge symmetry breakdown. 
The origin of such breaking can be a quantum mechanism similar to a 
Coleman - Weinberg mechanism \cite{coleman1} in $\lambda \phi^4$-theory. The 
essential difference is that in our case the symmetry breakdown mechanism 
should be non-perturbative one and its derivation now is the very big problem 
because we have not any analytical non-perturbative technique for the quantum 
field theories with strong interactions.
\par 
It is interesting to note that the regular solution for the colored flux tube 
exists only for a discrete spectrum of $m^*$ and $\phi^*$ parameters. 
Probably it is an indication of quantum origin of this object like to a 
discrete spectrum of energy levels of a quantum particle in a potential 
well. 
\par 
One can underline the difference between the Nielsen - Olesen and colored 
flux tubes: the first one is a topological solution but the another one 
is a pure dynamical object without any topological properties. 
\par 
Finally, the main result of this letter is that the confinement problem 
is closely connected with the problem of the symmetry breakdown: if symmetry 
breaking takes place then a colored flux tube solution exists. The 
mathematical problems here is the absense of any analytical technique 
for non-perturbative calculations in gauge field theories. It is 
possible that such technique can be based on the Heisenberg idea about 
the quantization of a non-linear spinor field \cite{heis}. 

\vspace{24pt}
\section{ACKNOWLEDZMENTS}
\vspace{12pt}

I am very grateful to the Alexander von Humboldt foundation for the 
financial support of this work and H.-J- Schmidt for the invitation to 
research in Potsdam University.

\end{document}